# In-Memory and Error-Immune Differential RRAM Implementation of Binarized Deep Neural Networks

M. Bocquet[1*], T. Hirztlin[2*], J.-O. Klein[2], E. Nowak[3], E. Vianello[3], J.-M. Portal[1] and D. Querlioz[2]

[1]Aix Marseille Univ, Université de Toulon, CNRS, IM2NP, Marseille, France
[2]C2N, Univ Paris-Sud, CNRS, Orsay, France, email: damien.querlioz@u-psud.fr
[3]CEA, LETI, Grenoble, France      *These authors contributed equally to the work

*Abstract*—RRAM-based in-Memory Computing is an exciting road for implementing highly energy efficient neural networks. This vision is however challenged by RRAM variability, as the efficient implementation of in-memory computing does not allow error correction. In this work, we fabricated and tested a differential $HfO_2$-based memory structure and its associated sense circuitry, which are ideal for in-memory computing. For the first time, we show that our approach achieves the same reliability benefits as error correction, but without any CMOS overhead. We show, also for the first time, that it can naturally implement Binarized Deep Neural Networks, a very recent development of Artificial Intelligence, with extreme energy efficiency, and that the system is fully satisfactory for image recognition applications. Finally, we evidence how the extra reliability provided by the differential memory allows programming the devices in low voltage conditions, where they feature high endurance of billions of cycles.

## I. INTRODUCTION

Deep neural networks are currently the most widely investigated architecture in Artificial Intelligence (AI) systems, with incredible achievements in image recognition, automatic translation, Go or Poker games. Unfortunately, when operated on central or graphics processing units (CPUs or GPUs), they consume considerable energy, in particular due to the intensive data exchanges between processors and memory [1,2]. Neural networks using in-memory computing (iMC) with RRAM are widely proposed as a solution to the Von-Neumann bottleneck [1]. However, RRAMs are prone to variability [3], and using Error Correcting Codes (ECC) as in more standard memories would ruin the benefits of iMC. ECCs indeed require large decoding circuits [4], which would need to be replicated multiple times in the case of iMC. This last point is the key challenge that we have to face for reliable neural networks on large RRAM memory arrays. In this paper, an experimental RRAM array with differential memory bit-cell (2T2R) based on $HfO_2$ devices, including all peripheral and a differential sensing scheme is fully characterized. This differential approach completely solves the key reliability challenge of large neural network implemented on RRAM using iMC concept. Due to its differential structure, our memory has intrinsically reduced errors. For the first time, we show that it improves reliability similarly to ECC with the same bit-cell count, whereas it considerably reduces CMOS overhead in the sensing scheme resulting in a clear gain in sensing speed. Additionally, we show that this structure allows the natural implementation of one of the most modern concepts of deep learning: Binarized Neural Networks (BNN) [5,6]. Such neural networks, can achieve state of the art AI performance, with very reduced memory requirements. Additionally, these networks use RRAMs as purely binary memories. We also show that the level of reliability achieved with our differential approach is fully appropriate as BNNs have an intrinsic tolerance to errors. We finally evidence that the robustness brought by our approach allows us to program RRAM devices at low voltage, where the devices feature very high endurance.

## II. DIFFERENTIAL MEMORY STRUCTURE: AN IDEAL ARCHITECTURE FOR IN-MEMORY COMPUTING

For this work, we fabricated memory arrays with a differential memory structure in a $HfO_2$-based OxRAM process, integrated in the BEOL of a 130 nm CMOS logic process [7], on top of the fourth metal layer (Cu) (Fig. 1). The OxRAM devices correspond to $TiN/HfO_2/Ti/TiN$ stacks. The thickness of both $HfO_2$ and Ti layers is 10 nm. Each bit is stored in a 2T2R structure in a complementary fashion: the two devices are programmed to complementary states (LRS/HRS or HRS/LRS) (Fig. 2). Each column features a differential precharge sense amplifier (PCSA) [8] (Fig. 2 and 3), which operates by comparing the resistance of the two memory devices. We fabricated and tested several structures with 2k devices, associated sense amplifiers and row and column decoders on chip. Fig. 4 first shows statistics of the forming process of the devices: all of them are formed, and the two devices do not influence each other. Fig. 5 shows the programming distribution in a low 55 µA Set compliance current ($I_c$) situation, prone to a high 1.2 % bit error rate in a 1T1R memory. Fig. 6 shows the response of all devices programmed in the same condition in a kbit 2T2R array as measured by our differential sense: only 0.2 % bit error is seen. Fig. 7 validates the functionality of the differential sense circuit with comprehensive testing. In previous works, 2T2R RRAM differential memories have already been fabricated, but their benefits on reliability have never been proven until now [9], [10], therefore we characterized our arrays extensively. Fig. 8 presents the mean number of bit errors on kbits array. We see that this number depends extensively on the programming conditions. Measurements with hundred millions of cycles on a single device, where the state of the devices is measured at each cycle (Fig. 9), evidence that LRS and HRS become less differentiated when the device ages, and that the 2T2R structure has much lower error rate than 1T1R in this situation. Overall, Fig. 10(a) shows that 2T2R always decreases the bit error rate with regards to 1T1R in diverse regimes. This Figure associates full array (device-to-device) measurements taken in a low compliance current regime, to address high error rates, and cycle-to-cycle experimental results in higher compliance current, to address lower error rates. The black curve is a theoretical result, assuming the PCSA has an ideal behavior, and therefore shows the minimum error rate achievable by the 2T2R approach. It is insightful to compare the benefits of 2T2R with the approach of ECC used in non-iMC contexts. Fig. 10(b) and Fig. 10(c) show the reliability benefits of various Single Error Correction (SEC) and Single Error Correction Double Error Detection (SECDED) codes. Interestingly, a code with the same memory redundancy as our approach ("SECDED(8,4)") leads to similar improvement in error rate. However, ECC decoding brings considerably more CMOS overhead than our approach: it needs logic circuitry to detect if an error occurred, and complex circuitry to detect the position of the error and correct it, requiring hundreds to thousands of logic gates [4]. This cost is unacceptable in iMC, as ECC decoders would need to be replicated for each memory array in the system, which can be hundreds. By contrast, our approach only uses a sense circuit that has no added complexity with regards to 1T1R solutions. Our work therefore extends the state of the art of RRAM iMC, where previous works do not propose a differential approach and are not compatible with technologies with errors [11,12]. In our approach, it is also possible to extend the sense amplifier to perform part of the logic, and thus to limit the CMOS overhead even further. For example, the circuit in Fig. 11 reads an RRAM cell, and at the same time performs a XNOR operation.

### III. USE OF THE DIFFERENTIAL ARRAY STRUCTURE FOR IMPLEMENTING DEEP NEURAL NETWORKS

Binarized Neural Networks (BNN) are ideally suited for exploiting our memory structure. They are conventional artificial neural networks, but weights and neuron activations are binary values instead of real numbers (Fig. 12). These systems require no multipliers, as they are replaced by XNOR logic operations, while additions are replaced by Popcount gates. BNNs can perform state of the art AI, with very reduced memory requirements [5,6]. This makes BNN ideal candidates for iMC. Fig. 13 shows how our memory structure can naturally implement iMC inference on BNN, associating a collection of kbit differential 2T2R memory arrays, all devices programmed in a binary fashion, with lightweight digital CMOS circuitry. XNOR operations can be performed directly in the PCSAs, or in separate logic gates. Popcount operations are based on 7-bits digital CMOS counters. Unlike most previous designs of neural networks with RRAM [13,14], this design is entirely digital, avoiding the need of high area operational amplifier or analog-to-digital converters. It does not require any multiplier, which allows extreme energy efficiency. We have designed the whole system based on synthesizable Verilog descriptions with Cadence IC design tools, and simulated it using the measured results on the memory arrays to model the memory blocks, and appropriate Value Change Dumps (VCD) inputs. This analysis was done with the design kit of a commercial 28 nm CMOS technology, to evaluate its potential on current technology. Fig. 14, which includes all CMOS overhead, highlights the amazing power efficiency of our design: it requires only nanoJoules to recognize one handwritten digit, while GPUs or CPUs-based AI requires micro to milliJoules.

### IV. ROBUSTNESS TO DEVICE VARIABILITY, POSSIBILITY TO USE THE DEVICES IN HIGH ENDURANCE REGIMES

We now investigate the impact of RRAM variability on an iMC BNN. We simulate our system for two tasks: handwritten character recognition task (MNIST) (Fig. 15ab), and a much more complicated photograph recognition task (CIFAR10), with a more complex deep neural network (Fig. 15cd). Without errors, our system can recognize 98.4% of the handwritten digits, and 87% of the photographs. Fig. 16 shows the impact of RRAM bit errors on the performance of the two tasks. Although errors change weight values between +1 and -1, up to $\sim 2\times 10^{-3}$ bit error rate can be tolerated with negligible impact on the performance of the neural network in both tasks. This is in contrast with most digital computing tasks, where the errors are catastrophic. The low demands of BNN in terms of errors, as well as the reliability brought by the 2T2R memory structure means that we can actually use RRAM devices in weak programming regimes where they individually are prone to errors. Fig. 17 shows that devices programmed in such weak conditions (reset voltage of 1.5V), with a 2T2R array bit error rate of $2\times 10^{-3}$, can show outstanding endurance of twenty billion cycles, that has very little impact on BNN performance.

### V. CONCLUSION

In this work, we showed experimentally that the 2T2R differential memory is a simple way to decrease the effect of RRAM variability, allowing comparable gains than SECDED error correction with a similar memory overhead, but without the associated area, time and energy overhead. The differential memory is also an ideal building block for in-memory BNN. We also showed that the relaxed requirements of BNNs in terms of errors, as well as the reliability benefit of the differential memory allows using RRAM devices in a low voltage regime, which implies extended endurance up to billions of cycles. These results highlight that although in-memory computing cannot rely on ECC, if a differential memory architecture is chosen, this does not have to translate into stringent requirements on device variability.

ACKNOWLEDGMENT

This work is supported by ERC grant NANOINFER (715872).

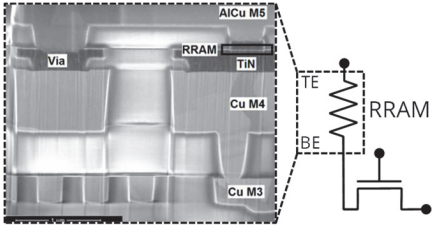

Fig. 1. (a) SEM cross-section of the TiN/HfO$_2$/Ti/TiN. Both HfO$_2$ and Ti layers are 10 nm thick. (b) Schematic view of the 1T1R cell configuration.

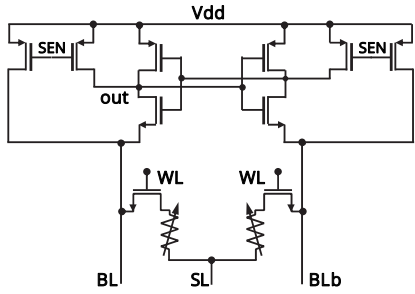

Fig. 2. Schematic of 2T2R precharge sense amplifier (PCSA).

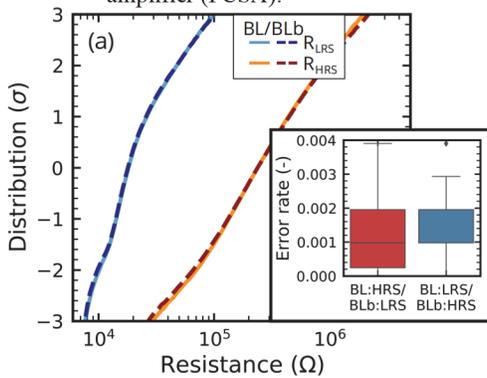

Fig. 3. (a) Photography and (b) schematic of the 2T2R array.

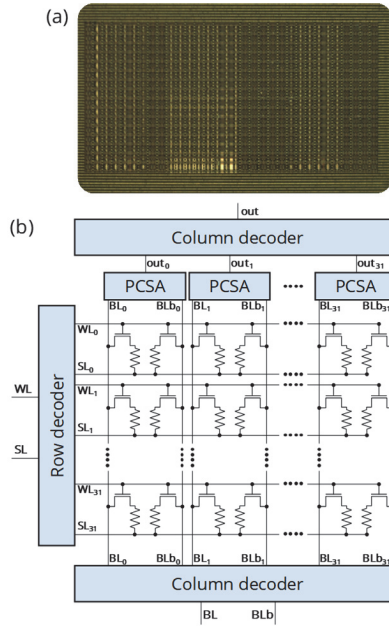

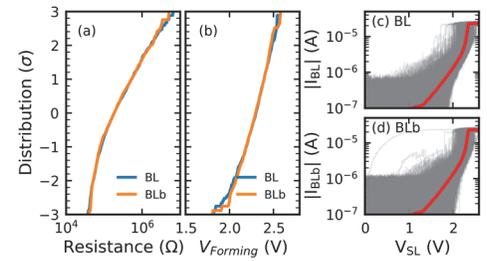

Fig. 4. Distribution of (a) resistance after forming and (b) forming voltages. I-V characteristics of forming operation for (c) BL cell and (d) BLb cell.

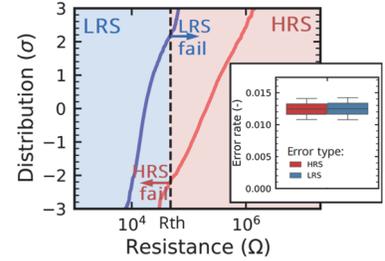

Fig. 5. Example of error rate extraction in 1T1R mode base on LRS/HRS distribution. Inset: bit error rate extracted. $V_{appReset}$=2.5V, $t_{Pulse}$=1μs, $I_c$=55μA.

Fig. 6. (a) Distribution of resistance for bit '1' (BL:HRS/BLb:LRS) and bit '0' (BL:LRS/BLb:HRS) in the case of a checkerboard type of programming the memory array. (b-c) Failure rate on 100 programming according differential sense for two checkerboards configuration. $V_{appReset}$=2.5V, $t_{Pulse}$=1μs, $I_c$=55μA.

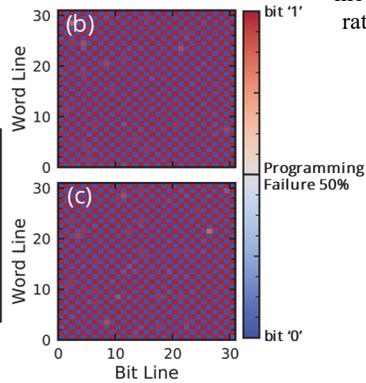

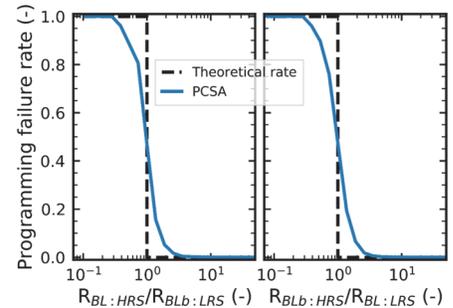

Fig. 7. Rate of programming failure indicated by the PCAS circuit as function of $R_{HRS}/R_{LRS}$ ratio obtained by a high-resolution resistance measurement @ Vread = 0.1V.

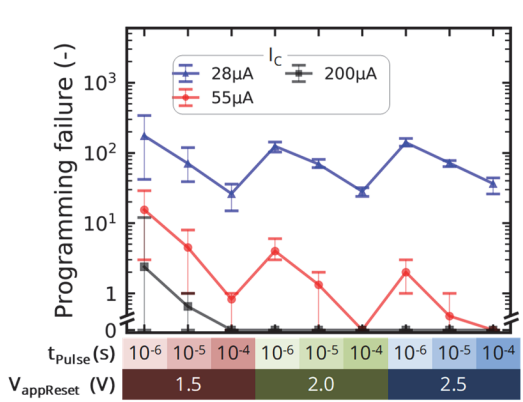

Fig. 8. Programming failure for different programming conditions for 2T2R configuration (PCSA) on a kbit 2T2R array.

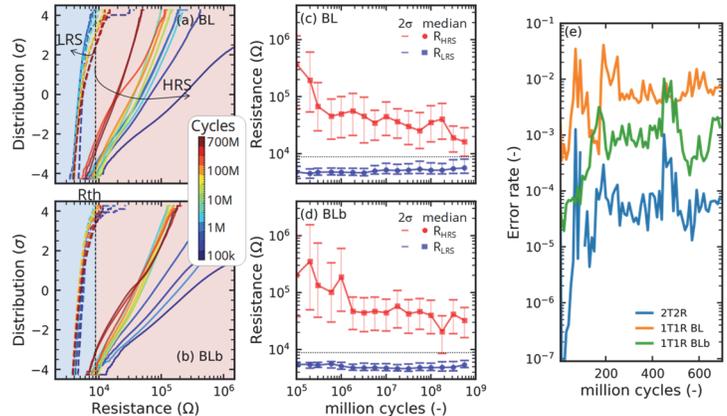

Fig. 9. a-b) The distribution of the resistance values, (c-d) the average value and (e) average error rate over 10 million cycles according to 2T2R configuration as function of number of cycles. $V_{appReset}$=2.5V, $t_{Pulse}$=1μs, $I_c$=200μA.

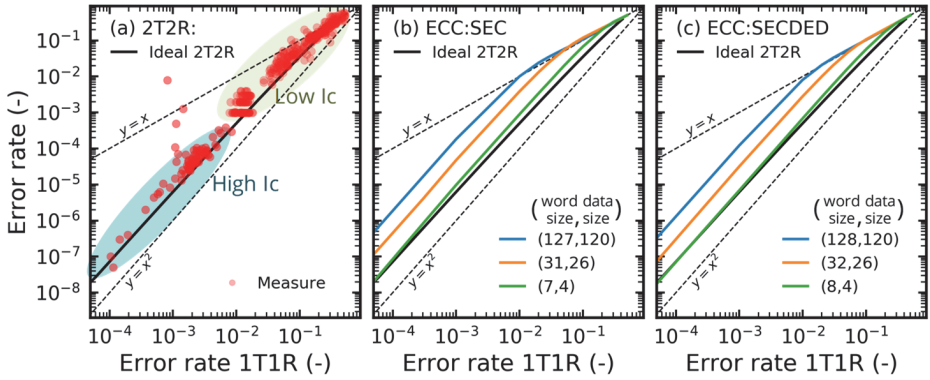
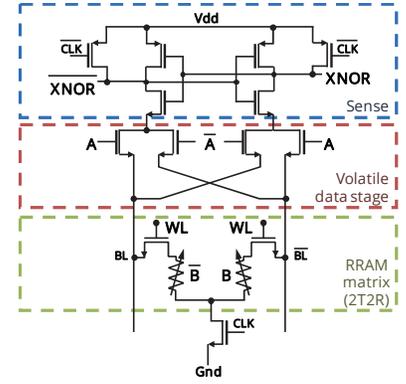

Fig. 10 (a) Experimental bit error rate of the 2T2R array as a function of the bit error rate on the individual (1T1R) RRAM devices. Bit error rate obtained with (b) SEC and (c) SECDED ECC as a function of the error rate on the individual devices.

Fig. 11. Adaptation of the PCSA circuit to perform a XNOR operation with the A input at the same time as READ operation.

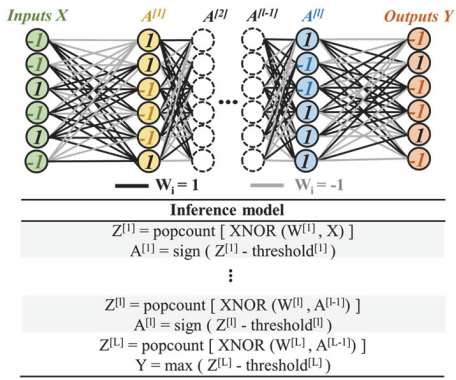
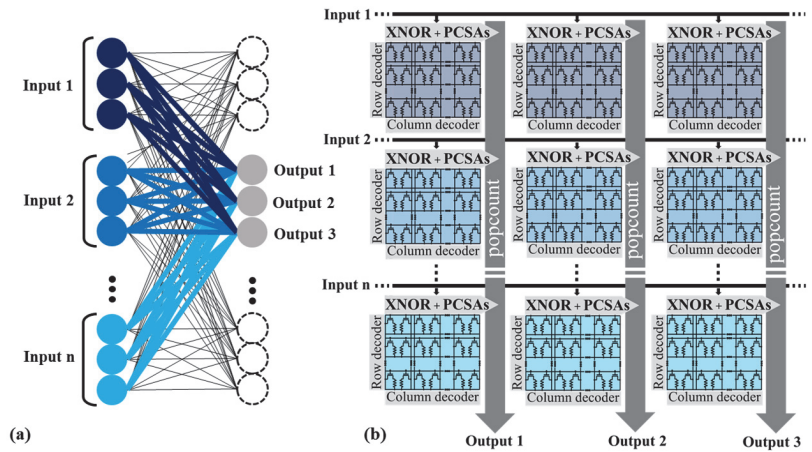

Fig. 12. Basic principle of a BNN. Synaptic weights $W$ and Neural Activations $A$ are binary values.

Fig. 13. Simplified architecture of an iMC BNN associating kbit 2T2R RRAM arrays with lightweight CMOS logic. The colors indicate the correspondence between formal neural network and hardware resources.

| RRAM In-memory BNN (this work) | 25nJ |
|---|---|
| RRAM In-memory 8-bit fixed point | 80nJ |
| Analog Phase Change Memory[*] [14] | ~56nJ |
| GPU (Tesla V100) | ~µJ |
| CPU (Xeon E5) | ~mJ |

Fig. 14. Comparison of the energy to recognize one handwritten digit, including all CMOS overhead. RRAM results are computed for a commercial 28 nm technology. [*]Taking into account inference-only, and scaled to the size of our neural network.

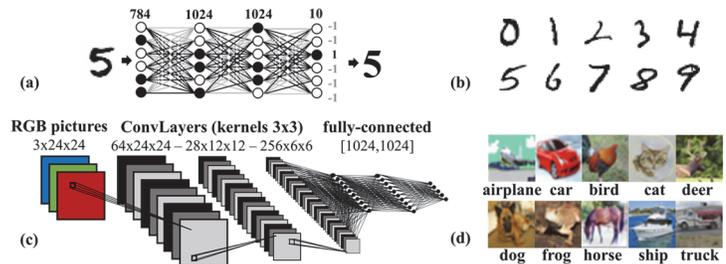

Fig. 15. Neural networks used for (a) digits recognition (MNIST) (c) photograph recognition (CIFAR10) – examples of digits (b) and photographs (d) to recognize.

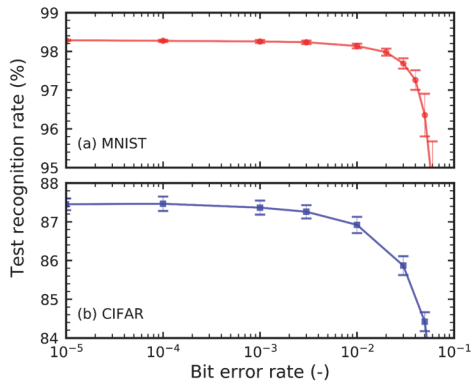

Fig 16. Dependence of the recognition rate of our BNN with the error rate of the memory arrays, for the MNIST and CIFAR10 tasks.

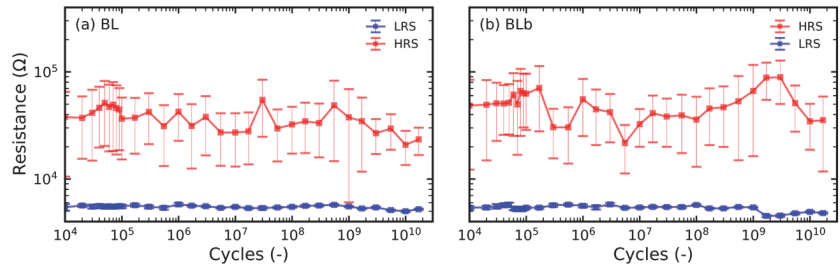

Fig. 17. Endurance measurement on two devices programmed at low voltage ($V_{RESET}$=1.5V), programming time of 1 µs and compliance current 200 µA. In this regime, the whole array 2T2R bit error rate is $2\times10^{-3}$.